\documentclass[fleqn,twoside]{article}
\usepackage{espcrc2}
\usepackage{epsfig}

\usepackage{graphicx}
\usepackage[figuresright]{rotating}

\def\det   {\mathop{\hbox{det}}}
\def\Det   {\mathop{\hbox{Det}}}

\def\Det   {\mathop{\hbox{Det}}}

\def\Staple{
\setlength{\unitlength}{0.9cm}
\begin{picture}(6.0,2.0)(4.5,0)
\thicklines
\put(0.0,0.1){\vector(0,1){0.9}}
\put(0.0,1.0){\vector(1,0){0.9}}
\put(0.9,1.0){\vector(0,-1){0.9}}
\end{picture}}

\def\Link{
\setlength{\unitlength}{0.9cm}
\begin{picture}(6.0,2.0)(0,0)
\thicklines
\put(0.0,0.0){\vector(1,0){0.9}}
\end{picture}}
\def\FiveStaple{
\setlength{\unitlength}{0.9cm}
\begin{picture}(6.0,2.0)(9.,0)
\thicklines
\put(0.0,0.1){\vector(0,1){0.9}}
\put(0.0,1.0){\vector(1,1){0.5}}
\put(0.5,1.5){\vector(1,0){0.9}}
\put(1.4,1.5){\vector(-1,-1){0.5}}
\put(0.9,1.0){\vector(0,-1){0.9}}
\end{picture}}

\def\SevenStaple{
\setlength{\unitlength}{0.9cm}
\begin{picture}(6.0,2.0)(13.5,0)
\thicklines
\put(0.0,0.1){\vector(0,1){0.9}}
\put(0.0,1.0){\vector(1,1){0.5}}
\put(0.5,1.5){\vector(1,2){0.3}}
\put(0.8,2.1){\vector(1,0){0.9}}
\put(1.7,2.1){\vector(-1,-2){0.3}}
\put(1.4,1.5){\vector(-1,-1){0.5}}
\put(0.9,1.0){\vector(0,-1){0.9}}
\end{picture}}

\makeatother

\newcommand{\AmS}{{\protect\the\textfont2
  A\kern-.1667em\lower.5ex\hbox{M}\kern-.125emS}}

\title{{ Dynamical Simulations with Smeared Link
Staggered Fermions 
} }

\author{A. Hasenfratz\address[MCSD]{Department of Physics, University of Colorado, 
        Campus Box 390, Boulder, CO 80309, USA}
        \thanks{Plenary talk presented at {\it Lattice 2002}, MIT, 
          Cambridge, USA, 24-29 June 2002.
        }
       }

\begin{document}

\begin{abstract}
\vspace{1pc}
One of the most serious problems of the staggered fermion lattice action
is flavor symmetry violation. Smeared link staggered fermions can
improve flavor symmetry by an order of magnitude relative to the standard
thin link action. Over the last few years different smearing transformations
have been proposed, both with perturbatively and non-perturbatively
determined coefficients. What hindered the acceptance and more general
use of smeared link fermions until now is the relative difficulty
of dynamical simulations and the lack of perturbative calculations
with these actions. In both areas there have been significant improvement
lately, that I will review in this paper.
\end{abstract}

\vskip -0.4cm
\maketitle

\section{INTRODUCTION}

Fermionic actions with smeared link gauge connections have gained
popularity in recent years. For staggered fermions, where the effect
of smearing on flavor symmetry restoration is easy to understand perturbatively
and straightforward to measure numerically, smearing became almost
mandatory \cite{Toussaint:2001zc,Bernard:2002bk,Hasenfratz:2002jn,Alexandru:2002jr}.
For Wilson-like fermions, including overlap and domain wall fermions
as well, the acceptance of smearing has been slower, though there
is evidence that smearing improves the chiral symmetry of Wilson fermions
and increases computational efficiency for all formulations \cite{DeGrand:1998mn,DeGrand:2000tf,Gattringer:2002xg,Gattringer:2002sb_lat02}.
The lack of interest in smeared link Wilson fermions is partially
due to the fact that there is no well defined physical quantity, like
flavor symmetry violation for staggered fermions, that would indicate
the effect of smearing for Wilson like fermions.  

Smearing can have undesirable effects as it can distort the (lattice)
short distance properties of the configurations \cite{Bernard:2000ht}.
What level this distortion is depends on the specific smearing and
is clearly observable form the short distance part of the heavy quark
potential. While excessive smearing is never desirable, for certain
operators, like heavy fermions, only very little or no smearing is
acceptable. Fortunately there is no need to use the same smearing
transformation, or even the same fermionic formulations, for heavy
and light quarks, so this problem is easily solvable. 

Smearing both for staggered and Wilson fermions can be considered
as part of a systematic improvement program where irrelevant operators
are added to the action to remove lattice artifacts. Perturbative
improvement systematically removes lattice corrections of order \( O(a^{n}) \)
and \( O(g^{n}) \), while the fixed point action program removes
lattice artifacts using a non-perturbative, renormalization group
based approach. At the end both methods modify the fermionic action
by including non-nearest neighbor quadratic and higher order terms
and modify the gauge connection by replacing the simple link by some
sort of extended paths. The systematic fixed point action program
has been used mainly with Wilson like fermions while the systematic
perturbative improvement program has been applied for staggered fermions,
but both methods can (and to some extent have been) extended to both
fermionic formulation. 

In this talk I will review some of the recent developments regarding
staggered fermions. The two smeared staggered actions that are used
in dynamical simulations at present are the so called Asqtad and HYP
actions. Since the Asqtad action has been reviewed in detail last
year \cite{Toussaint:2001zc}, I will concentrate on the newer HYP
action \cite{Hasenfratz:2001hp}. Since the HYP action contains SU(3)
projected gauge links, it cannot be simulated with the usual molecular
dynamics based algorithms. I will review an updating method that is
based on the well-known pseudo-fermion update and is efficient for
smeared link fermions \cite{Hasenfratz:2002jn,Hasenfratz:2002ym,Hasenfratz:2002pt}.
Finally I show some preliminary results obtained with this algorithm
for HYP staggered fermions.

\section{THE ADVANTAGES OF SMEARED LINK ACTIONS}

I start with a brief outline of the perturbative and non-perturbative
methods of constructing smeared link staggered fermion actions. The
perturbative method is systematic, in principle it can be extended
to arbitrary order in \( O(a^{n}) \) and \( O(g^{n}) \), though
at present only a fully \( O(a^{2}) \) improved action, the Asqtad
action, has been used in numerical simulations \cite{Lepage:1998vj,Orginos:1999cr,Trottier:2001gy}.
The non-perturbative method I will review is the hypercubic smearing
(HYP) transformation \cite{Hasenfratz:2001hp}. At present it is not
part of a systematic improvement program, yet it is a smearing transformation
that appears to be about a factor of two better in flavor symmetry
restoration than the Asqtad action and incidentally it is also \( O(a^{2}) \)
improved. 

\textbf{Perturbative smeared link actions:} Staggered fermions break
flavor symmetry as gluons at the edge of the Brillouin zone connect
fermions with different flavors. The tree level perturbative improvement
program removes the flavor changing gluons by requiring that the smeared
gauge link \( B_{\mu } \) vanishes at momenta \( (\pi ,0,0,0) \),
\( (\pi ,\pi ,0,0) \), etc. This can be achieved by the addition
of three different kind of gauge loops of length three, five, and
seven, connecting nearest neighbor fermions \[
\Link \Staple \FiveStaple \SevenStaple \, .\]
 This smearing is frequently referred to as Fat7 \cite{Orginos:1999cr}. The
construction cannot remove all flavor changing gluons, those with
momentum near but not exactly \( \pi  \) do not vanish, but their
coefficients are reduced. Removing the smeared gluons at the corners
of the Brillouin zone introduces an \( O(a^{2}) \) term in the small
momentum gluon. This term can be removed by the addition of yet another
gauge term, the so called Lepage term \cite{Lepage:1998vj}. To make
the action fully \( O(a^{2}) \) improved the pure gauge part of the
action is chosen to be the tadpole improved 1-loop Symanzik action
and the Dirac operator is modified by the addition of a third neighbor
Naik term, constructed with the original thin links, leading to the
Asqtad action. 

At tree level perturbation theory SU(3) projected (unitarized) links
cannot be distinguished from the non-projected ones. By choice, in
the Asqtad action the extended gauge loop connections appear linearly,
no projection to SU(3) is included. The Asqtad action has been studied
extensively by the MILC collaboration and many of its properties have
been reviewed last year in \cite{Toussaint:2001zc} . 

Recently a more ambitious program of 1-loop improvement has been developed
and preliminary results were reported at this conference. The 1-loop,
\( O(a^{2}g^{2}) \) improved action contains about a dozen new four
fermion coupling terms, some with imaginary coefficients. In numerical
simulations the four fermion terms can be resolved by the introduction
of auxiliary scalar fields, but the presence of imaginary couplings
could create more problems. To date the action has not been used neither
in quenched nor in dynamical simulations. Most of the work so far
concentrated on identifying the different terms of the action, determining
their coefficients and estimating their importance. It would make
future numerical simulations easier if only a few of the theoretically
necessary terms were important, especially if the imaginary terms
could be neglected. The analysis of the perturbative action therefore
is very important.  

\textbf{Non-perturbative smeared link actions:} To design a non-perturbative
improvement scheme first one has to understand the origin of flavor
symmetry violation non-perturbatively. Staggered fermions break flavor
symmetry because each Dirac and flavor component of the fermion field
occupies a different lattice site of the hypercube, and couples to
different gauge fields. If the gauge links within a hypercube fluctuate
strongly, the different fermionic components observe very different
gauge environment and act like independent fields, and not like components
of a Dirac spinor, thus causing flavor symmetry violation. This very
simple physical picture suggests that in order to improve the flavor
symmetry of staggered fermions one has to minimize the gauge field
fluctuations at the hypercubic or plaquette level. More precisely,
one should minimize the tail of the plaquette distribution, i.e. maximize
the smallest plaquettes of the configurations, as these are the ones
causing most of the flavor symmetry violations. In a quenched study
with different smearing transformations the relation between the minimum
plaquette distribution and flavor symmetry violation was clearly demonstrated
suggesting a fast, non-perturbative method to optimize smearing \cite{Hasenfratz:2001hp}.

HYP smearing was constructed to be very local and optimized through
the minimum plaquette distribution. It is a succession of three levels
of modified SU(3) projected APE smearing steps, each level modified
such that at the end none of the original links extend beyond the
hypercubes that attach to the smeared link

{\noindent \centering {\small \begin{eqnarray*}
V_{i,\mu } & = & Proj_{SU(3)}[(1-\alpha _{1})U_{i,\mu }\\
 &  & +\frac{\alpha _{1}}{6}\sum _{\pm \nu \neq \mu }\tilde{V}_{i,\nu ;\mu }\tilde{V}_{i+\hat{\nu },\mu ;\nu }\tilde{V}_{i+\hat{\mu },\nu ;\mu }^{\dagger }]\, ,
\end{eqnarray*}
 \begin{eqnarray}
\tilde{V}_{i,\mu ;\nu } & = & Proj_{SU(3)}[(1-\alpha _{2})U_{i,\mu }\label{hyp_smear} \\
 &  & +\frac{\alpha _{2}}{4}\sum _{\pm \rho \neq \nu ,\mu }\bar{V}N_{i,\rho ;\nu \, \mu }\bar{V}_{i+\hat{\rho },\mu ;\rho \, \nu }\bar{V}_{i+\hat{\mu },\rho ;\nu \, \mu }^{\dagger }]\, \nonumber 
\end{eqnarray}
}\small \par}

{\noindent \centering {\small \begin{eqnarray*}
\bar{V}_{i,\mu ;\nu \, \rho } & = & Proj_{SU(3)}[(1-\alpha _{3})U_{i,\mu }\\
 &  & +\frac{\alpha _{3}}{2}\sum _{\pm \eta \neq \rho ,\nu ,\mu }U_{i,\eta }U_{i+\hat{\eta },\mu }U_{i+\hat{\mu },\eta }^{\dagger }]\, .
\end{eqnarray*}
}\small \par}

In Eq. \ref{hyp_smear} \( V_{i,\mu } \) denotes the HYP smeared
links, \( U_{i,\mu } \) are the original thin links, while \( \tilde{V}_{i,\mu } \)
and \( \bar{V}_{i,\mu } \) are intermediate smeared links. There
are three parameters in HYP smearing, the \( \alpha _{i} \) APE parameters
of each level. They can be optimized non-perturbatively by minimizing
fluctuations at the plaquette level. One can also optimize the parameters
of the HYP links perturbatively by requiring that the smeared gauge
links \( B_{\mu } \) vanish at the edges of the Brillouin zone. The
perturbative condition leads to parameters that are, at least after
tadpole improvement, consistent with the non-perturbatively optimized
parameter values. In numerical simulations the non-perturbatively
optimized parameters \begin{equation}
\alpha _{1}=0.75,\; \alpha _{2}=0.60,\; \alpha _{3}=0.35
\end{equation}
are used. 

HYP smeared links are fairly local. That is best seen in the measurement
of the static quark potential. The static potentials measured with
thin gauge links and with HYP smeared links agree, up to an additive
constant, at distances \( r/a\geq 2 \) \cite{Hasenfratz:2001tw}.
The difference between the potential values is significant only at
\( r/a=1 \), and even there it can be well described by the perturbative
lattice Coulomb potential. An added bonus of the HYP potential is
that its statistical errors are about an order of magnitude smaller
than the thin link potential errors, making the measurement of the
potential parameters, especially the string tension, feasible even
on relatively small data sets.

The tail of the plaquette distribution that is maximized in HYP smearing
contains mainly those plaquettes, frequently called dislocations,
that are responsible for the residual chiral symmetry breaking in
domain wall fermions, and for the large real eigenmodes that slow
down the overlap fermion computations. A smeared link action that
is optimized to remove these modes will likely improve domain wall
and overlap computations as well \cite{DeGrand:2000tf}.

\subsection{Flavor Symmetry Restoration in Smeared Link Actions}

\begin{figure}
\vskip -0.2cm
{\centering \resizebox*{8cm}{!}{\rotatebox{-90}{\includegraphics{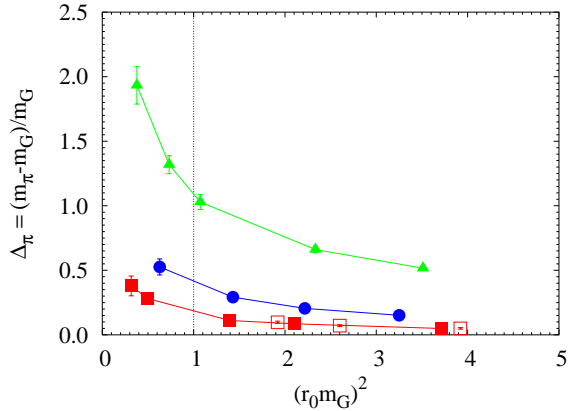}}} \par}

\vskip -0.4cm
\caption{Flavor symmetry violation with the standard thin (triangles), Asqtad
(circles) and HYP (filled squares) staggered fermions on \protect\( a=0.17\protect \)fm
quenched configurations. The open squares correspond to dynamical
HYP data at similar lattice spacing. The vertical line at \protect\( (r_{0}m_{G})^{2}=1\protect \)
correspond to \protect\( m_{G}\approx 400\protect \)MeV pions. 
\label{delta_pi}}
\vskip -0.4cm
\end{figure}

The effect of smearing on flavor symmetry is illustrated in Figure
\ref{delta_pi} where the relative split between the Goldstone pion
\( m_{G} \) and lightest non-Goldstone pion \( m_{\pi } \) \begin{equation}
\Delta _{\pi }=\frac{m_{\pi }-m_{G}}{m_{G}}
\end{equation}
is plotted as the function of the square of the Goldstone pion mass
measured in units of \( r_{0} \) \cite{Sommer:1994ce}. The triangles,
circles and filled squares in Figure \ref{delta_pi} are results obtained
with the standard thin link, Asqtad and HYP fermion actions on \( 8^{3}\times 24 \),
\( \beta =5.7 \) Wilson plaquette gauge configurations with lattice
spacing \( a\approx 0.17 \)fm. The open squares correspond to dynamical
HYP configurations at similar lattice spacing. They agree with the
quenched HYP data, indicating that improvement seen in quenched simulations
carry over to dynamical ones. For all mass values the HYP action has
about an order of magnitude smaller flavor symmetry violations than
the standard thin link action, and it is about a factor of two better
than Asqtad fermions. It is an interesting question why HYP fermions
have better flavor symmetry restoration than Asqtad fermions. Both
are perturbative tree level improved, but the HYP links have three
levels of SU(3) projections and more gauge loops at each level. Whether
it is the SU(3) projection, the extra gauge loops, or both, that make
HYP fermions better is not obvious. Recently it has been shown \cite{Lee:2002fj}
that actions with projected SU(3) links that have the same tree level
\( O(a^{2}) \) improvement are also equivalent at 1-loop level. If
that perturbative statement holds true non-perturbatively as well,
one could argue that the SU(3) projection is the key to the improved
flavor symmetry of the HYP action. The non-perturbative assumption
can be easily tested either by quenched measurements of flavor symmetry
with projected Asqtad/Fat7 actions or by simply measuring the distribution
of the tail of the plaquette with these actions.

\subsection{Perturbative Properties of Smeared Actions }

A frequent criticism toward smeared (or, for that matter, any kind
of improved) actions is that their perturbative properties are not
known. This excuse has always been a weak one, and now, that several
papers have been published showing that perturbative calculations
with smeared links, even projected ones, are doable, the criticism
now longer holds. Rather, one should ask if the perturbative properties
of smeared actions are better or worse than the standard thin link
actions. One can recall that with thin link actions, both with staggered
and Wilson fermions, the 1-loop perturbative matching factors for
most physically interesting quantities are large. That makes the connection
between the lattice and continuum schemes like \( \overline{MS} \)
unreliable. It has been observed that in the case of staggered fermions
the perturbative contribution from the fermion doublers is as large
as from the usual gluon tadpoles \cite{Golterman:1998jj}. Since smearing
removes the fermion doublers, it should also improve the perturbative
matching factors. Several recent works showed that this is indeed
what happens \cite{Trottier:2001gy,Lee:2002bf,Lee:2002fj,Lee:2002ui}.
For example in \cite{Lee:2002ui}  the 1-loop matching coefficients
for several staggered bilinear operators were calculated for different
smeared actions. These matching coefficients connect continuum to
lattice as \begin{eqnarray}
O^{\rm {cont}}_{i} & \! \! =\! \!  & O_{i}^{\rm {latt}}\big (1+\frac{C_{F}g^{2}}{16\pi ^{2}}(2d_{i}ln(\mu a)+c_{i,i})\big )\\
 &  & +\rm {off-diagonal\quad terms}.\nonumber 
\end{eqnarray}
 If a coefficient \( c_{i,i}\! \approx 5 \) at \( 1/a=2 \)GeV lattice
spacing, the 1-loop perturbative corrections are about 10\%. Table
\ref{c00_coeff} shows the scalar-scalar coefficient that is relevant
for the fermionic condensate for the standard thin link, Asqtad and
HYP actions. Obviously both the Asqtad and HYP actions satisfy the
minimum requirement \( c_{1,1}<5 \), their matching factors are 10-30
times smaller than the thin link value. 
\begin{table}
\begin{tabular}{|c|c|c|c|}
\hline 
\( \!  \)action&
\( \!  \)Thin link&
\( \!  \)Asqtad&
\( \!  \)HYP\\
\hline
\hline 
\( c_{1,1} \)&
-29.3&
-2.2&
-0.6\\
\hline
\end{tabular}

\vskip .5cm

\caption{The perturbative matching factors for the scalar-scalar bilinear
for the standard thin link, Asqtad and HYP fermions \cite{Lee:2002bf,Lee:2002ui}\label{c00_coeff}}
\vskip -0.2cm
\end{table}
 The HYP value is even small enough to use on rougher, \( 1/a=1.2 \)GeV
configurations where one finds \begin{equation}
\label{sig_cont}
\Sigma (2GeV)_{\overline{MS}}=\frac{1}{Z_{P/S}}\Sigma _{\rm {latt}}(1/a=1.2\rm {GeV}),
\end{equation}
 with \begin{equation}
\label{Z_ps}
Z_{P/S}=0.92.
\end{equation}
 This perturbatively determined renormalization Z factor agrees with
the value obtained via the non-perturbative matching method of \cite{Hernandez:2001yn}.
I will discuss that further in Sect. 4.

\section{DYNAMICAL SIMULATIONS WITH SMEARED ACTIONS}

Smeared actions that are linear in the thin gauge links, like the
Asqtad action, can be simulated with the usual molecular dynamics
based algorithms. Calculating the fermionic force for these actions,
while tedious, is straightforward. The only problem observed with
these simulations is the unusually large autocorrelation time for
the topological charge \cite{Bernard:2002sa}. 

Actions with projected smeared links, like the HYP action, can not
be simulated using molecular dynamics methods as the fermionic force
is prohibitively expensive to evaluate. However these actions are
smooth enough in the gauge fields to be simulated using a modified
version of the well known pseudo-fermion algorithm, the partial-global
stochastic metropolis or PGSM update \cite{Hasenfratz:2002jn,Alexandru:2002jr,Hasenfratz:2002ym,Hasenfratz:2002pt}.
The PGSM algorithm satisfies the detailed balance condition \cite{Grady:1985fs},
and with smeared link fermions it can be an efficient algorithm. 

We consider a smeared link action of the form\begin{equation}
\label{full_action}
S=S_{g}(U)+\bar{S}_{g}(V)+S_{f}(V),
\end{equation}
 where \( S_{g}(U) \) and \( \bar{S}_{g}(V) \) are pure gauge actions
depending on the thin links \( \{U\} \) and smeared links \( \{V\} \),
respectively, and \( S_{f} \) is the fermionic action depending on
the smeared links only.The smeared links are constructed deterministically,
like hypercubic (HYP) blocking. Any gauge action can be used for \( S_{g}(U) \)
while \( \bar{S}_{g}(V) \) is chosen to optimize the algorithm. The
fermionic action describing \( n_{f} \) degenerate flavors of staggered
fermions is \begin{equation}
\label{s_f_orig}
S_{f}(V)=-\frac{n_{f}}{4}\ln \, \Det \, \Omega (V)
\end{equation}
 with \( \Omega =M^{\dagger }M \) defined on even sites only. In
the PGSM update first a subset of the thin links \( \{U\} \) are
updated and a new thin gauge link configuration \( \{U'\} \) is proposed
with transition probability that satisfies detailed balance with \( S_{g}(U) \).
The proposed configuration is accepted with the probability \begin{equation}
\label{Pstoch}
P_{\rm {acc}}=\rm {min}\{1,e^{-\Delta \bar{S}_{g}}e^{-\xi ^{*}[A-1]\xi }\},
\end{equation}
 where \( \Delta \bar{S}_{g}=\bar{S}_{g}(V^{'})-\bar{S}_{g}(V) \)
is the difference in the smeared gauge actions, \( A=\Omega ^{'-n_{f}/8}\Omega ^{n_{f}/4}\Omega ^{'-n_{f}/8} \),
and the stochastic vector \( \xi  \) is generated with Gaussian distribution.
In Eq. \ref{Pstoch} only one stochastic vector is used in each update
step, the acceptance probability approximates the ratio of the old
and new actions \begin{equation}
e^{-\Delta \bar{S}_{g}}\Det \, ^{n_{f}/4}\Omega ^{'}\Omega ^{-1}=e^{-\Delta \bar{S}_{g}}<e^{-\xi ^{*}[A-1]\xi }>
\end{equation}
only on average. If the stochastic formula estimates this ratio poorly,
the autocorrelation time of the update will be large, it could even
be infinite. This can be seen if we consider the standard deviation
of the stochastic estimator \begin{equation}
\det \, ^{-1}(A)=<e^{-\xi ^{*}[A-1]\xi }>_{\xi },
\end{equation}
\begin{eqnarray}
\sigma _{A}^{2} & = & <e^{-2\xi ^{*}[A-1]\xi }>-<e^{-\xi ^{*}[A-1]\xi }>^{2}\nonumber \\
 & = & \det \, ^{-1}(2A-1)-\det \, ^{-2}(A).\label{standard_dev} 
\end{eqnarray}
 Eq. \ref{standard_dev} is valid only if the matrix \( 2A-1 \) is
positive definite. If the matrix \( A \) has even one eigenvalue
that is less than or equal to 1/2, the formula in Eq. \ref{standard_dev}
is not valid, the standard deviation is infinite. Unfortunately this
is frequently the case as the eigenvalues of the fermionic matrices
\( \Omega  \) and \( \Omega ^{'} \) vary from \( 1/4m^{2} \) to
above 16. 

There are several ways to improve the efficiency of the stochastic
estimator \cite{Alexandru:2002jr}. Here I mention only one, the determinant
breakup. If we rewrite the fermionic action of Eq. \ref{s_f_orig}
as \begin{equation}
S_{f}(V)=-\ln \, \Det \, ^{n_{f}n_{b}/4}\, \Omega (V)^{1/n_{b}}
\end{equation}
with \( n_{b} \) an arbitrary integer, the fermionic determinant
can be estimated using \( n_{b}n_{f}/4 \) random source vectors as
\begin{equation}
\label{Pstoch_many}
\det \, ^{-1}(A)=<e^{-\sum ^{n_{b}n_{f}/4}_{i=1}\xi _{i}^{*}[\tilde{A}-1]\xi _{i}}>_{\xi _{i}}
\end{equation}
 with \( \tilde{A}=\Omega ^{'-n_{f}/8n_{b}}\Omega ^{n_{f}/4n_{b}}\Omega ^{'-n_{f}/8n_{b}} \).
Now the standard deviation is finite as long as none of the eigenvalues
of \( A \) is less than \( 1/2^{n_{b}} \), a condition that is much
easier to satisfy. In fact for every finite quark mass it is possible
to choose \( n_{b} \) such that the condition is satisfied. The determinant
breakup reduces the statistical fluctuations of the stochastic estimator
by taking the sum of \( n_{b}n_{f}/4 \) smaller terms instead of
the original term\begin{equation}
-\xi ^{*}[A-1]\xi \longrightarrow -\sum ^{n_{b}n_{f}/4}_{i=1}\xi _{i}^{*}[\tilde{A}-1]\xi _{i}.
\end{equation}
 While taking the average of several stochastic estimates in the acceptance
step of the original PGSM algorithm (Eq. \ref{Pstoch}) would violate
the detailed balance condition, the determinant breakup procedure
is still exact.

An added bonus of the determinant breakup is that now simulating arbitrary
number of degenerate or non-degenerate flavors is straightforward
as long as \( n_{b} \) is chosen to be a multiple of 4.

With the determinant breakup of Eq. \ref{Pstoch_many} the acceptance
probability can be close to the theoretical maximum, \( e^{-\Delta \bar{S}_{g}}\Det ^{n_{f}/4}\Omega ^{'}\Omega ^{-1} \).
The PGSM update will be effective only if this value is close to one,
i.e. the proposed configurations \( \{U\}\longrightarrow \{U'\} \)
in the first step of the algorithm are close to the dynamical configurations
of \( S \). Fairly good agreement can be achieved if we choose \( S_{g}(U) \)
such that it matches the average plaquette and/or correlation length
of the dynamical configurations. That requires the shift of the pure
gauge coupling that can be compensated by the second pure gauge action
term, \( \bar{S}_{g}(V) \). Since this term depends on the smeared
links, its fluctuations are greatly suppressed and including this
term in the acceptance rate does not change it significantly.

\begin{figure}
\vskip -0.4cm
{\centering \resizebox*{8cm}{!}{\rotatebox{-90}{\includegraphics{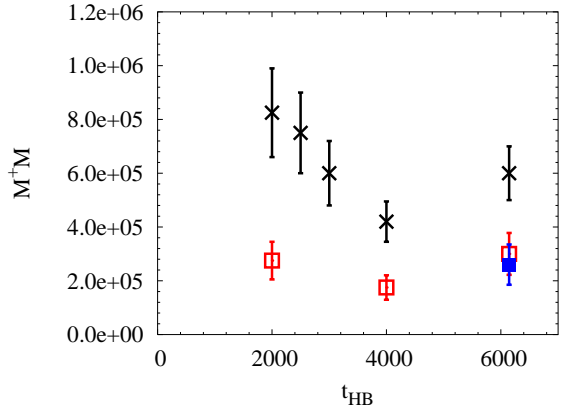}}} \par}

\vskip -0.4cm
\caption{Cost of creating independent configurations with \protect\( n_{f}=2\protect \)
flavor HYP action, measured in fermionic matrix multiplies. Results
are from 10fm\protect\( ^{4}\protect \) lattices with lattice spacing
\protect\( a\approx 0.17\protect \)fm. Open squares: \protect\( n_{b}=8\protect \),
\protect\( m_{\pi }/m_{\rho }\approx 0.7\protect \), filled square:
\protect\( n_{b}=12\protect \), \protect\( m_{\pi }/m_{\rho }\approx 0.7\protect \),
crosses: \protect\( n_{b}=12\protect \), \protect\( m_{\pi }/m_{\rho }\approx 0.55\protect \)
simulations. \label{M+M_cost}}
\vskip -0.4cm
\end{figure}

The fractional powers of the fermionic matrices in \( \tilde{A} \)
can be evaluated using polynomial approximation \cite{Montvay:1997vh,Alexandru:2002sw},
but to evaluate the acceptance probability according to Eq. \ref{Pstoch_many}
requires \( n_{b}n_{f}/4 \) multiplication with \( \tilde{A} \),
a considerable computational overhead. A detailed study of the autocorrelation
of the PGSM update can be found in \cite{Alexandru:2002jr}. Figure
\ref{M+M_cost} shows the cost, measured in \( M^{\dagger }M \) matrix
multiplies, of creating statistically independent configurations that
are separated by two autocorrelation time lengths as the function
of the number of links changed in \( \{U\} \) to create \( \{U^{'}\} \)
in the first step of the PGSM algorithm. The data was obtained with
\( n_{f}=2 \) flavors of dynamical HYP fermions on \( 8^{3}\times 24, \)
about 10fm\( ^{4} \) volume configurations with different determinant
breakup parameters (\( n_{b}=8,12 \)) and quark masses corresponding
to \( m_{\pi }/m_{\rho }\approx 0.7 \) and 0.55. In case of the heavier
quark mass the optimal update touches about 4000 links or around 10\%
of the total links on this volume. With an acceptance rate of 50\%,
\( 2\times 10^{5} \) \( M^{\dagger }M \) multiplies are needed to
create statistically independent configurations. For comparison, creating
independent configurations with the standard thin link staggered action
at similar lattice spacing at \( m_{\pi }/m_{\rho }\approx 0.7 \)
would require about \( 6-8\times 10^{5} \) matrix multiplies, a few
times more than with the HYP action using the PGSM algorithm. 

The PGSM algorithm is a volume square algorithm. Creating independent
configurations on larger volumes will become more and more expensive.
However, judging from Figure \ref{M+M_cost}, the computational cost
of creating even 100-200fm\( ^{4} \) configurations is not more than
5-10 times more expensive than molecular dynamics update of thin link
staggered fermions.

\section{SIMULATION RESULTS WITH HYP FERMIONS}

The results summarized in this section are preliminary, obtained on
modest size lattices at large lattice spacing with limited computational
resources. They are in no way comparable to the extensive data collected
with the Asqtad action over the last couple of years, some of it reported
at this conference \cite{Bernard:2002bk,Bernard:2002yd}. My goal
in this section is to show that the PGSM method is applicable, and
at least for standard quantities, nothing unexpected is seen with
the HYP action.

Simulations with the HYP action were performed on \( 8^{3}\times 24 \)
lattices with \( a\approx 0.17 \)fm lattice spacing with three quark
masses corresponding to roughly \( m_{\pi }/m_{\rho }=0.7, \) 0.65
and 0.55. To monitor finite size effects and to study the scaling
of the algorithm the simulations were repeated on \( 12^{3}\times 36 \)
lattices with the two smaller quark mass values. While no autocorrelation
time measurement is available on the larger lattices, other factors
indicate that the algorithm scales with the square of the volume,
maybe even better when comparing the small \( 8^{3}\times 24 \) lattices
to the larger volume. 

\begin{figure}
\vskip -0.4cm
{\centering \resizebox*{8cm}{!}{\rotatebox{-90}{\includegraphics{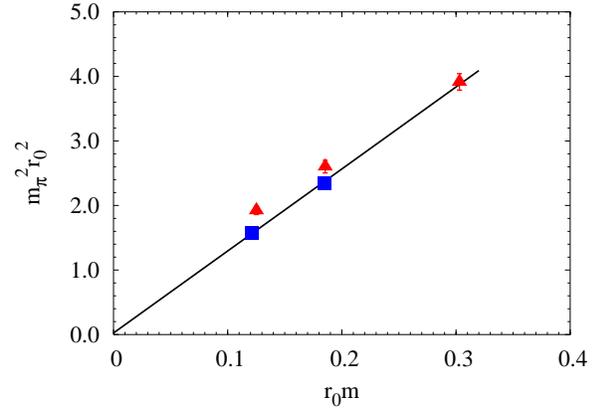}}} \par}

\vskip -0.4cm
\caption{The Goldstone pion mass squared as the function of the quark mass,
in units of \protect\( r_{0}\protect \). Triangles: \protect\( 8^{3}\times 24\protect \),
squares: \protect\( 12^{3}\times 36\protect \) lattices, all with
\protect\( r_{0}\sim 3.0\protect \).\label{mpi_vs_mq}}
\vskip -0.4cm
\end{figure}

In Figure \ref{delta_pi} I have already showed that flavor symmetry
restoration with the dynamical HYP action is consistent with the quenched
result. Figure \ref{mpi_vs_mq} shows the scaling of the Goldstone
pion mass squared with the quark mass, both measured in units of \( r_{0} \).
The large finite volume distortion on the smaller volumes is surprising
at first, quenched simulations with similar \( m_{\pi }/m_{\rho } \)
ratios and volumes do not show such effect. However the data is actually
consistent with finite volume chiral perturbative calculations \cite{Sharpe:2001fh,Bernard:2001yj}.
According to these calculations finite volume corrections of the data
above are expected to be a few percent on the smaller volume with
the largest quark mass and also on the larger volume with the smallest
quark mass, larger for the other data points on the smaller volume.
Using data with small finite volume corrections we can estimate the
renormalization factor \( Z_{P/S} \) based on the non-perturbative
method of \cite{Hernandez:2001yn}. The value we obtain is consistent
with the perturbative prediction in Sect. 2.2 \begin{equation}
Z_{P/s}=0.92(1)\qquad \rm {(non-perturbative)}.
\end{equation}

\begin{figure}
\vskip -0.4cm
{\centering \resizebox*{8cm}{!}{\rotatebox{-90}{\includegraphics{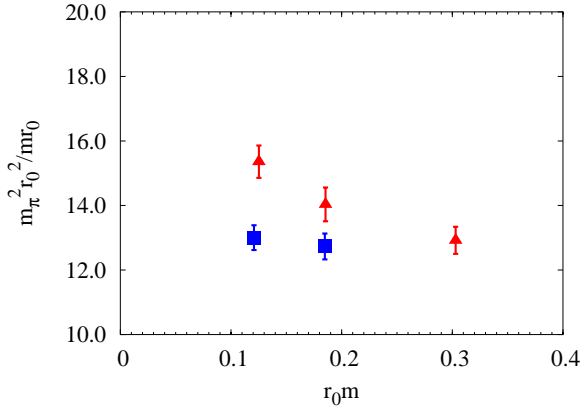}}} \par}

\vskip -0.4cm
\caption{\protect\( m^{2}_{\pi }/m\protect \) as the function of the quark
mass, in units of \protect\( r_{0}\protect \). Notation is the same
as in Figure \ref{mpi_vs_mq}. \label{mpiomq_vs_mq}}
\vskip -0.4cm
\end{figure}

 Can we see chiral logarithms in this data? Figure \ref{mpiomq_vs_mq}
shows the ratio \( m^{2}_{\pi }/m_{q} \) as the function of the quark
mass, again in units of \( r_{0}. \) The data points with small finite
volume corrections are consistent with a constant value, at these
quark masses there is no obvious sign for chiral logarithms. We can
use this data to predict the chiral condensate \( \Sigma  \) through
the GMOR relation. With the renormalization factor obtained above
the chiral condensate, translated to \( \overline{MS} \) scheme at 2GeV is
predicted to be\begin{equation}
\label{sigma_ms_value}
\Sigma _{\overline{MS}}(2GeV)r^{3}_{0}=0.36(4).
\end{equation}
 This value, though with large errors, seems to be larger than the
quenched value \( \Sigma ^{Quen}_{\overline{MS}}(2GeV)r^{3}_{0}=0.30(4) \)
predicted by several groups using a much more sophisticated finite
volume method that can be applied only with exactly chiral fermions
\cite{DeGrand:2001ie,Hernandez:2001yn,Hasenfratz:2002rp}. The difference
between the quenched value and Eq. \ref{sigma_ms_value} could be
only numerical accuracy, though one would expect a larger condensate
from dynamical simulations. New results at smaller lattice spacing
and larger volumes can resolve this question in the near future.

\begin{figure}
\vskip -0.4cm
{\centering \resizebox*{8cm}{!}{\rotatebox{-90}{\includegraphics{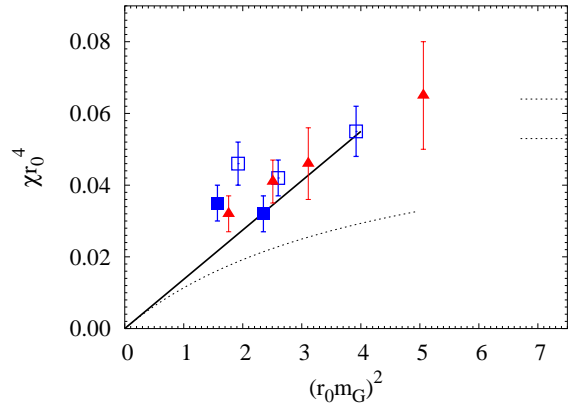}}} \par}

\vskip -0.4cm
\caption{The topological susceptibility as the function of the Goldstone pion
mass squared, in units of \protect\( r_{0}.\protect \) Triangles:
standard thin link staggered fermions at \protect\( a\approx 0.10\protect \)fm
\cite{Hasenfratz:2001wd}; open squares: HYP action on \protect\( 8^{3}\times 24\protect \)
lattices, \protect\( a\approx 0.17\protect \)fm; filled squares:
HYP action on \protect\( 12^{3}\times 24\protect \) lattices, \protect\( a\approx 0.17\protect \)fm.
The dashed curve is the prediction from \cite{Durr:2001ty}, the short
lines on the right indicate the quenched value.\label{suscept_plot} }
\vskip -0.4cm
\end{figure}
I conclude this section with some recent results on the topological
susceptibility with HYP fermions. At small quark masses the topological
susceptibility is expected to scale linearly with the square of the
Goldstone pion mass with a coefficient that is predictable from chiral
perturbation theory (solid line in Figure \ref{suscept_plot}). Higher
order corrections might push the value even lower (dashed curve in
Figure \ref{suscept_plot}) \cite{Durr:2001ty}. Numerical simulations
have a hard time reproducing this behavior, especially at small quark
masses and large lattice spacing, even with improved fermions \cite{Hasenfratz:2001wd,Bernard:2002sa}.
In Figure \ref{suscept_plot} I show results that were obtained with
thin link staggered fermions at \( a\approx 0.1 \)fm lattice spacing
and with HYP fermions at \( a\approx 0.17 \)fm lattice spacing. The
topological charge was measured after 2-3 levels of HYP smearing using
an improved topological charge operator \cite{DeGrand:1997gu}. The
difference between the smaller and larger volume results with HYP
fermions is mainly due to the change of \( r_{0} \) and \( m_{\pi } \)
due to finite volume effects. The topological susceptibility is quite
independent of the volume. While even HYP fermions do not reproduce
the theoretical expectations, they seem to be consistent with the
thin link action results at 70\% smaller lattice spacing. It is unlikely
that the difference between the numerical results and theoretical
expectations are due to the remnant flavor symmetry violations. Even
the heaviest of the non-Goldstone pions would predict a smaller topological
susceptibility than the measured value at the smallest quark mass.
The resolution to this problem might come from exploring the difference
of what objects the fermions identify as instantons versus what objects
the gluonic charge operators identify as instantons in these simulations.

\section{CONCLUSION}

Smeared gauge links remove short scale vacuum fluctuations, improving
the flavor symmetry of staggered fermion actions. In addition to improved
flavor symmetry, smeared fermion actions have improved perturbative
properties, like 10-30 times reduced 1-loop matching factors. While
these actions are more complicated than thin link ones, with the newly
developed partial-global stochastic Metropolis (PGSM) update even
SU(3) projected actions can be simulated on fairly large volumes.
There are already extensive numerical data with the dynamical Asqtad
action using the molecular dynamics based R algorithm. The HYP action
has been used in exploratory studies only with the PGSM algorithm. 

Many of the properties of the smeared staggered actions should apply
to Wilson like fermions as well. Smearing could improve chiral symmetry,
and reduce the occurrence of exceptional configurations. Smeared actions
also require smaller clover coefficients. In domain wall fermions
smearing can reduce the residual chiral symmetry breaking while for
overlap fermions smearing increases computational efficiency. These
properties are worth exploring in the future.

\bibliographystyle{h-elsevier}
\bibliography{lattice}

\end{document}